\documentclass[%
 aip,
 amsmath,amssymb,
 reprint,%
]{revtex4-1}
\usepackage{xcolor}
\usepackage{graphicx}% Include figure files
\usepackage{dcolumn}% Align table columns on decimal point
\usepackage{bm}% bold math
\usepackage{ulem}
\usepackage[utf8]{inputenc}
\usepackage[T1]{fontenc}
\usepackage{mathptmx}
\usepackage{bm}
\usepackage{etoolbox}

\makeatletter
\def\@email#1#2{%
 \endgroup
 \patchcmd{\titleblock@produce}
  {\frontmatter@RRAPformat}
  {\frontmatter@RRAPformat{\produce@RRAP{*#1\href{mailto:#2}{#2}}}\frontmatter@RRAPformat}
  {}{}
}%
\makeatother
\begin{document}

\preprint{AIP/123-QED}

\title{Continuous-wave AC magnetometry up to hundreds of megahertz using nitrogen-vacancy centers under a strong transverse magnetic field}
% Force line breaks with \\
\author{Ryusei Okaniwa}
\affiliation{\mbox{School of Fundamental Science and Technology, Keio University, Yokohama, Kanagawa 223-8522, Japan}}
\affiliation{\mbox{Center for Spintronics Research Network, Keio University, Yokohama, Kanagawa 223-8522, Japan}}
\author{Rui Suzuki}
\affiliation{\mbox{School of Fundamental Science and Technology, Keio University, Yokohama, Kanagawa 223-8522, Japan}}
\affiliation{\mbox{Center for Spintronics Research Network, Keio University, Yokohama, Kanagawa 223-8522, Japan}}
\author{Yuichiro Matsuzaki}%
\affiliation{Department of Electrical, Electronic, and Communication Engineering, Faculty of Advanced Science and Engineering, Chuo University, Kasuga, Bunkyo-ku, Tokyo 112-8551, Japan}
%\\This line break forced with \textbackslash\textbackslash%
\author{Norio Tokuda}
\affiliation{\mbox{Advanced Research Center for Diamond Science and Technology, Kanazawa University, Kanazawa, Ishikawa, 920-1192, Japan}}
\author{Junko Ishi-Hayase}
\email[Author to whom correspondence should be addressed: ]{hayase@appi.keio.ac.jp}
\affiliation{\mbox{School of Fundamental Science and Technology, Keio University, Yokohama, Kanagawa 223-8522, Japan}}
\affiliation{\mbox{Center for Spintronics Research Network, Keio University, Yokohama, Kanagawa 223-8522, Japan}}
\homepage{http://www.Second.institution.edu/~Charlie.Author.}

\date{\today}% It is always \today, today,

\begin{abstract}
Nitrogen-vacancy (NV) centers in diamond are promising quantum sensors for AC magnetic fields. However, conventional dynamical-decoupling-based magnetometry becomes technically demanding at high frequencies because of the requirement for increasingly short and high-power microwave pulses. This paper describes an AC magnetometry technique based on continuous-wave optically detected magnetic resonance (CW-ODMR) measurements, which enables the detection of signals in the hundred-MHz range. By applying a strong static magnetic field transverse to the NV axis, the energy separation between the relevant electron-spin states was tuned and resonantly coupled with the target radiofrequency (RF) magnetic field, thereby forming RF-dressed states. Unlike pulsed dynamical-decoupling schemes, this approach does not require high-power microwave pulse sequences. The key feature of our method is that the detectable frequency can be tuned by controlling the bias magnetic field strength. Using this methodology, we have experimentally achieved the frequency-tunable detection of AC magnetic fields from 14.4--632~MHz. The sensitivity was 0.558~$\mu{\rm T}/\sqrt{\rm Hz}$ at 14.4 MHz and 24.4~$\mu{\rm T}/\sqrt{\rm Hz}$ at 632 MHz. We characterized the field-dependent ODMR contrast and linewidth, and found that their degradation at large transverse fields is a important factor limiting the high-frequency sensitivity. These results establish a simple CW-ODMR approach for widely tunable RF magnetic-field sensing without high-power pulsed spin control.
\end{abstract}

\maketitle

Nitrogen-vacancy (NV) centers in diamond have emerged as a leading platform for quantum information,\cite{bradley2019ten, guo2024single} communication,\cite{humphreys2018deterministic, pompili2021realization} and sensing.\cite{maze2008nanoscale, zhou2025entanglement}  Featuring atomic-scale dimensions and long coherence times, even at room temperature,\cite{balasubramanian2009ultralong, herbschleb2019ultra} NV centers enable high-sensitivity, high-spatial-resolution detection of diverse physical quantities, including magnetic fields,\cite{taylor2008high, barry2024sensitive} electric fields,\cite{dolde2011electric, cheng2023radio} and temperature.\cite{kucsko2013nanometre, tabuchi2023temperature} These superior properties have driven extensive applications across device diagnostics,\cite{garsi2024three} materials characterization,\cite{nishimura2023wide} and biomedicine.\cite{yu2024noninvasive}

NV magnetometry generally requires laser excitation for initialization and readout of the NV electron spin state,\cite{harrison2004optica} alongside microwave (MW) irradiation for coherent NV spin control.\cite{jelezko2004observation} For AC sensing in the kHz--MHz range, Hahn echo and dynamical decoupling techniques, such as the Carr-Purcell-Meiboom-Gill (CPMG) method\cite{pham2012enhanced} and the XY family,\cite{kehayias2017solution} are widely used to extend the coherence time and improve magnetic field sensitivity. However, limited MW power typically constrains the detectable bandwidth to around 30~MHz.\cite{jung2025impedance} 

To approach the tens-to-hundreds of MHz regime, methods such as continuous dynamical decoupling,\cite{stark2017narrow, hermann2024extending, ding2025wide} the AC Zeeman effect,\cite{li2019wideband}  quantum frequency mixing,\cite{wang2022sensing, karlson2024quantum, yin2025high} and conventional Rabi magnetometry\cite{alsid2023solid} have been developed. These approaches have substantially extended the accessible frequency range; however, many of them require pulsed or multi-tone control. A simple continuous-wave (CW) method for detecting AC magnetic fields over a broadly tunable frequency range is therefore highly desirable. Although a CW method for AC magnetic-field sensing has previously been demonstrated,\cite{magaletti2022quantum} it detects field components transverse to the NV quantization axis. 

In this paper, we propose a CW optically detected magnetic resonance (ODMR)-based sensing methodology that can measure AC fields along the NV axis of up to several hundred MHz. CW-ODMR avoids complex MW pulse timing and pulse-sequence control. Additionally, the CW readout scheme is potentially compatible with wide-field imaging using CCD/CMOS cameras at a slow frame rate. The key concept is radiofrequency (RF)-dressed states with a strong transverse magnetic field. The RF-dressed states are produced by the strong coupling between the NV electron spin and RF waves. While our group previously achieved AC sensing via RF-dressed states,\cite{saijo2018ac, yamaguchi2019bandwidth,dokai2026broadband} the detection range was limited to a few MHz by the intrinsic diamond strain. Applying a strong transverse magnetic field makes the RF-interacting energy levels tunable, resolving this issue. In this study, we extended our previous RF-dressed-states sensing scheme by applying a strong transverse bias magnetic field. The transverse field continuously tunes the energy separation between the RF-coupled NV spin states, thereby extending the detectable AC-field frequency far beyond the strain-limited regime of our previous work. We report the results of experiments that demonstrate CW-ODMR-based AC magnetometry from 14.4--632~MHz and investigate the field dependence of the ODMR contrast and linewidth, which determine the achievable sensitivity at high frequencies.

We briefly outline our proposed method. Supplementary material Sec.~S1 contains detailed theoretical derivations.

Unlike conventional NV magnetometry using an axial static magnetic field, our method applies a bias magnetic field $B_x$ perpendicular to the NV axis. 
The total Hamiltonian \(\hat{H}_0\) is given by
\begin{align}
\label{eq:1}
\hat{H}_0=D\hat{S}_z^2 + E_x(\hat{S}_x^2 -\hat{S}_y^2)+\gamma_eB_x\hat{S}_x,
\end{align}
where \(\hat{S}_i\) (\(i = x, y, z\)) is the spin-1 operator of the NV electron spin, \(D/2\pi\) (\(\simeq2.87\)~GHz) is the zero-field splitting, \(E_x/2\pi\) is the intrinsic strain along the \(x\) direction, and \(\gamma_e/2\pi\) (\(\simeq28\)~GHz/T) is the gyromagnetic ratio of the electron spin. 
Under the weak transverse field, the eigenstates are approximately \{\(|+\rangle, |-\rangle,|0\rangle\)\}, where \(|\pm\rangle\) can be described by \(|\pm\rangle=(|+1\rangle\pm|-1\rangle)/\sqrt{2}\).\cite{zanelli2025interplay} The energy separation %\( E_x'\)
between \(|+\rangle\) and \(|-\rangle\) can be written as follows:
\begin{align}
\label{eq:2}
2E_x'\simeq 2E_x+\frac{(\gamma_eB_x)^2}{D+E_x},
\end{align}
where $B_x$ plays the same role as the intrinsic strain under $|\gamma_eB_x| \ll D$.

\begin{figure}[ht]
\centering
\includegraphics[width=8.5cm]{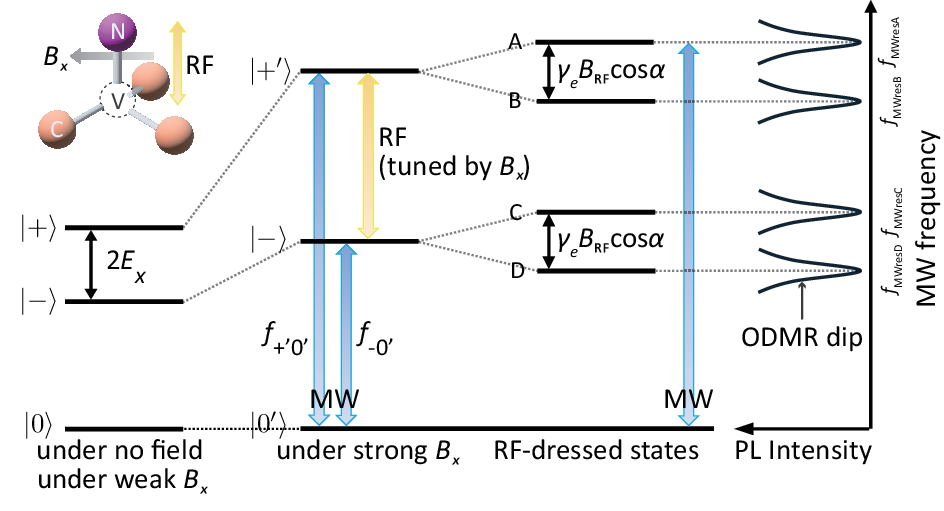}
\caption{Energy level diagram of the NV electron spin in the ground state manifold. Under no field or a weak transverse bias field, the energy eigenstates are approximately \(\{|+\rangle\), \(|-\rangle\), \(|0\rangle\)\} (\(|\pm\rangle=(|+1\rangle \pm|-1\rangle)/\sqrt{2}\)). By applying a stronger bias field, these eigenstates evolve to \(\{|+'\rangle\), \(|-\rangle\), \(|0'\rangle\)\}. \(|+'\rangle\) and \(|0'\rangle\) are defined in Eq. \eqref{eq:3}. This transverse magnetic field increases both the MW transition frequencies and the energy separation between $\vert +'\rangle$ and $\vert -\rangle$. A target AC field resonant with this separation is tunable by $B_x$ and induces a coherent interaction and creates the RF-dressed states (labeled A--D). In the CW-ODMR measurements, four photoluminescence (PL) dips are expected to appear.}
\setlength\abovecaptionskip{3pt}
\label{NVstate}
\end{figure}

Under a strong transverse magnetic field, this approximation no longer holds, necessitating new eigenstates and energy levels, as shown in Fig.~\ref{NVstate}.\cite{gao2026dressed} First, because $\hat{S}_x = |+\rangle\langle 0| + |0\rangle\langle +|$, $B_x$ mixes $|+\rangle$ and $|0\rangle$, forming the superpositions $|+'\rangle$ and $|0'\rangle$. The transformed basis satisfies the following relations:\cite{matsuzaki2025theoretical} 
\begin{align}
\label{eq:3}
|+\rangle=\cos\alpha |+'\rangle-\sin\alpha |0'\rangle,\ |0\rangle=\sin\alpha |+'\rangle+\cos\alpha |0'\rangle,
\end{align}
where the parameter \(\alpha\) is determined by and increases with $B_x$. 

Second, the MW transition frequencies shift as the transverse field increases. These frequencies \(f_{+'0'}\) (\(|+'\rangle \Leftrightarrow |0'\rangle\)) and \(f_{-0'}\) (\(|-\rangle \Leftrightarrow |0'\rangle\)) can be written as
\begin{align}
2\pi f_{+'0'}&=\sqrt{(D+E_x)^2+(2\gamma_eB_x)^2},\label{eq:4}\\
2\pi f_{-0'}&=\frac{1}{2}D-\frac{3}{2}E_x+\frac{1}{2}\sqrt{(D+E_x)^2+(2\gamma_eB_x)^2}.\label{eq:5}
\end{align}
Both frequencies shift upward with increasing $B_x$,\cite{lamba2024vector} and the RF-driven transition frequency (\(2\pi(f_{+'0'}-f_{-0'})\)) changes accordingly. The corresponding energy separation is given by
\begin{align*}
\label{eq:6}
2\pi(f_{+'0'}-f_{-0'})&=-\frac{1}{2}D+\frac{3}{2}E_x+\frac{1}{2}\sqrt{(D+E_x)^2+(2\gamma_eB_x)^2}.
\stepcounter{equation}\tag{\theequation}
\end{align*}

Unlike our previous work driving the $|+\rangle \Leftrightarrow |-\rangle$ transition, we apply an RF field at frequency $2\pi(f_{+'0'} - f_{-0'})$, resonant with $|+'\rangle \Leftrightarrow |-\rangle$, to generate the RF-dressed states. Crucially, Eq.~\eqref{eq:6} shows that this resonant frequency is $B_x$-tunable. As shown in Fig.~\ref{NVstate}, RF dressing forms four energy levels (labeled A, B, C, and D). Applying the rotating wave approximation in an appropriate rotating frame yields the corresponding MW transition frequencies:
\begin{align}
\label{eq:7}
2\pi f_{\mathrm{MWresA}}=\sqrt{(D+E_x)^2+(2\gamma_eB_x)^2}+\frac{\gamma_eB_{\mathrm{RF}}}{2}\cos\alpha,
\end{align}
\begin{align}
\label{eq:8}
2\pi f_{\mathrm{MWresB}}=\sqrt{(D+E_x)^2+(2\gamma_eB_x)^2}-\frac{\gamma_eB_{\mathrm{RF}}}{2}\cos\alpha,
\end{align}
\begin{align}
\label{eq:9}
2\pi f_{\mathrm{MWresC}}=\frac{1}{2}D-\frac{3}{2}E_x&+\frac{1}{2}\sqrt{(D+E_x)^2+(2\gamma_eB_x)^2}  \notag  \\
&+\frac{\gamma_eB_{\mathrm{RF}}}{2}\cos\alpha,
\end{align}
\begin{align}
\label{eq:10}
2\pi f_{\mathrm{MWresD}}=\frac{1}{2}D-\frac{3}{2}E_x&+\frac{1}{2}\sqrt{(D+E_x)^2+(2\gamma_eB_x)^2}  \notag  \\
&-\frac{\gamma_eB_{\mathrm{RF}}}{2}\cos\alpha.
\end{align}
On resonance, the separation between each pair of RF-dressed MW resonances is therefore proportional to $B_{\rm RF}\cos\alpha$. This relation provides the basis for estimating the target AC magnetic-field amplitude from the CW-ODMR response. In CW-ODMR measurements, the RF-dressed states appear as split dips in the CW-ODMR spectrum. The target RF frequency can be tuned by varying $B_x$ because the \(|+'\rangle \Leftrightarrow |-\rangle\) energy separation depends on the transverse bias field. The detectable RF frequency range is broadened by $B_x$, with its lower limit set by the zero-field strain $2E_x$ ($B_x = 0$).

The experimental setup is described below. We used the same diamond sample as in Refs. \onlinecite{tabuchi2023temperature, okaniwa2024frequency}. The ensemble NV centers were preferentially aligned along the [111] direction and were contained in a nitrogen-doped CVD-grown layer in the (111) plane with an estimated thickness of 1.5~$\mu$m. The coherence time ($T_2$) and the dephasing time ($T_2^*$) were determined to be 3.3~$\mu$s and 231~ns, respectively. A transverse magnetic field along the $x$ direction was generated by a pair of neodymium magnets mounted on electric actuators, allowing field orientation and strength control by adjusting the magnet distance. MW fields were applied using an antenna beneath the diamond,\cite{sasaki2016broadband} whereas the target RF magnetic field was generated by a copper wire placed on the diamond surface. The NV photoluminescence (PL) was collected using a confocal microscope and detected with an avalanche photodiode. For additional details on the basic experimental configuration, see the supplementary material (Sec. S2).

\begin{figure*}[ht]
\centering
\includegraphics[width=17cm]{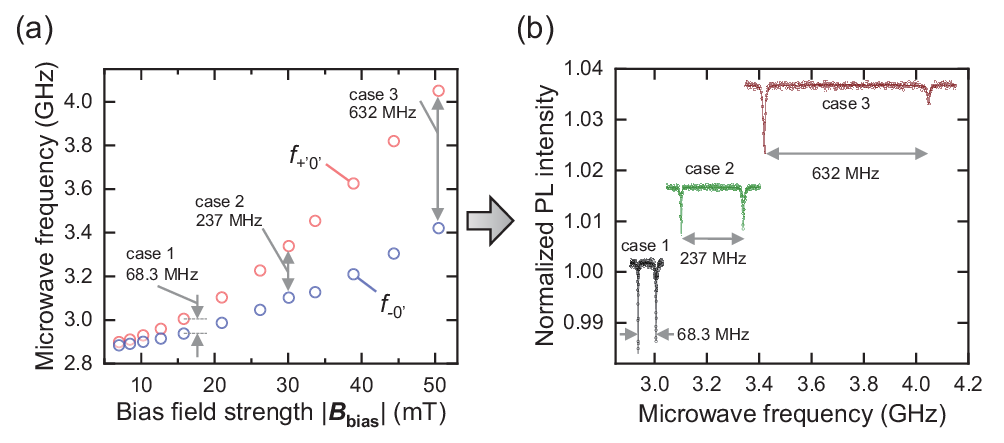}
\caption{(a) MW transition frequencies as a function of $\lvert\bm{B}_{\mathrm{bias}}\rvert$. Each $\lvert\bm{B}_{\mathrm{bias}}\rvert$ is estimated from ODMR spectra. Red and blue dots indicate the higher ($f_{+'0'}$) and lower ($f_{-0'}$) transitions, respectively. Both shift to higher frequencies with increasing $\lvert\bm{B}_{\mathrm{bias}}\rvert$. 
(b) CW-ODMR spectra at $\lvert\bm{B}_{\mathrm{bias}}\rvert = 15.8, 30.4$, and $50.7\text{~mT}$ for cases 1--3, corresponding to the pairs of dots enclosed by wavy boxes in (a). ODMR spectra for cases 2 and 3 are vertically offset for clarity. Spectra for cases 1 and 2 were measured at MW power of $-30\text{~dBm}$, whereas that for case 3 was recorded at $-20\text{~dBm}$ to enhance the PL contrast.}
\setlength\abovecaptionskip{3pt}
\label{result1}
\end{figure*}

The experimental results are summarized below. Figure~\ref{result1}(a) shows the MW transition frequencies \(f_{+'0'}\) and \(f_{-0'}\) as a function of the bias magnetic-field strength \(\lvert \bm{B}_{\mathrm{bias}}\rvert\), determined by Lorentzian fitting of the CW-ODMR spectra, as illustrated in Fig.~\ref{result1}(b). Both the resonance frequencies and the energy splitting between \(\lvert +'\rangle\) and \(\lvert -\rangle\) increase with \(\lvert \bm{B}_{\mathrm{bias}}\rvert\). By varying the transverse field, we experimentally tuned the \(|+'\rangle \Leftrightarrow |-\rangle\) transition frequency from 14.4~MHz to 632~MHz. 

To account for an axial component \(B_z\) caused by magnet misalignment and the geomagnetic field, we estimated \(B_x\) and \(B_z\) by diagonalizing the Hamiltonian and used gradient descent to minimize the squared deviations between the frequencies extracted from the experiment and those obtained by diagonalization. The resulting bias field magnitude \(\lvert \bm{B}_{\mathrm{bias}}\rvert=\sqrt{B_x^2+B_z^2}\) is plotted on the horizontal axis of Fig.~\ref{result1}(a). The transverse component dominates the fraction of \(\lvert \bm{B}_{\mathrm{bias}}\rvert\) over the investigated field range. For more details, see the supplementary material (Sec. S3).

\begin{figure*}[ht]
\centering
\includegraphics[width=16cm]{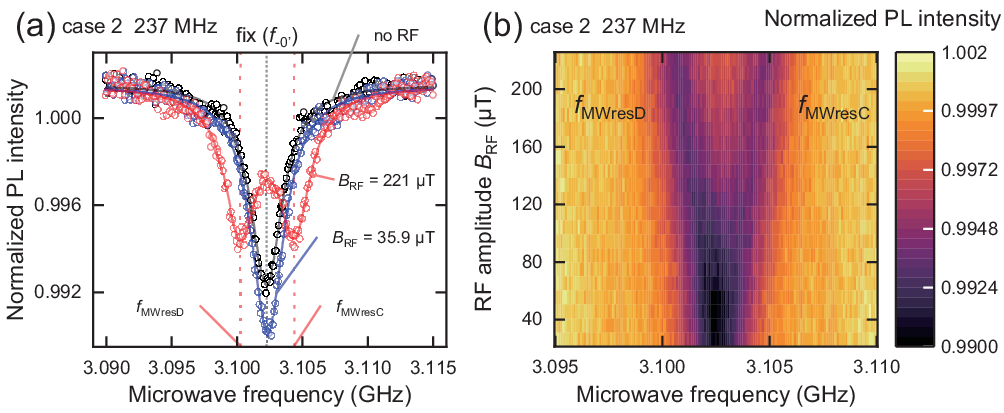}
\caption{(a) CW-ODMR spectra with Lorentzian fitting for various RF amplitudes at $\vert{}\bm{B}_{\mathrm{bias}}\vert{} = 30.4\text{~mT}$ (case 2), focusing on the $f_{-0'}$ transition. The black curve corresponds to the case without RF, showing a single PL dip. The red and blue curves represent data at $B_{\mathrm{RF}} = 221~\mu\text{T}$ and $35.9~\mu\text{T}$, respectively. The red curve exhibits two distinct PL dips ($f_{\mathrm{MWresC}}$ and $f_{\mathrm{MWresD}}$ described in Eqs. \eqref{eq:9}-- \eqref{eq:10}), indicated by pink dotted lines, signifying the formation of the RF-dressed states. The PL intensity was recorded at a fixed MW frequency $f_{-0'}$, as indicated by the gray dotted line. (b) Colormap of CW-ODMR spectra as a function of the RF amplitude, where darker regions correspond to PL dips. The energy splitting of the RF-dressed states increases monotonically with RF amplitude.}
\setlength\abovecaptionskip{3pt}
\label{RFdependence}
\end{figure*}

Next, we verified the formation of the RF-dressed states at a representative frequency in the hundred-MHz range. Figure~\ref{RFdependence}(a) shows the RF-induced CW-ODMR spectrum modification around the transition \(f_{-0'}\), where $\lvert \bm{B}_{\mathrm{bias}}\rvert$ is 30.4~mT and the energy separation between \(|+'\rangle\) and \(|-\rangle\) is approximately 237~MHz. Under RF irradiation (\(B_{\mathrm{RF}}=221\)~\(\mu\)T), two distinct dips appear as red dots in Fig.~\ref{RFdependence}(a), confirming the formation of RF-dressed states. Figure~\ref{RFdependence}(b) displays the RF strength dependence of these dips. Their energy gap expands monotonically with increasing RF power. Consequently, RF sensitivity can be determined from the RF-amplitude-dependent change in the normalized PL intensity at a fixed MW frequency. We used the $f_{-0^\prime}$ transition for sensing because it exhibited a narrower linewidth (weaker first-order susceptibility to transverse-field fluctuations) than $f_{+^\prime0^\prime}$, as discussed below. As described next, signals were successfully detected throughout the investigated frequency range, demonstrating widely frequency-tunable AC magnetometry.

To determine RF sensitivity, we require normalized signal standard deviation \(\delta S\) per second and the signal derivative \(\mathrm{d}S/\mathrm{d}B_{\mathrm{RF}}\), similar to Refs. \onlinecite{saijo2018ac, yamaguchi2019bandwidth}. Here, $S$ is change in normalized PL intensity. Here, we focus on \(\mathrm{d}S/\mathrm{d}B_{\mathrm{RF}}\). 

\begin{figure*}[ht]
\centering
\includegraphics[width=17cm]{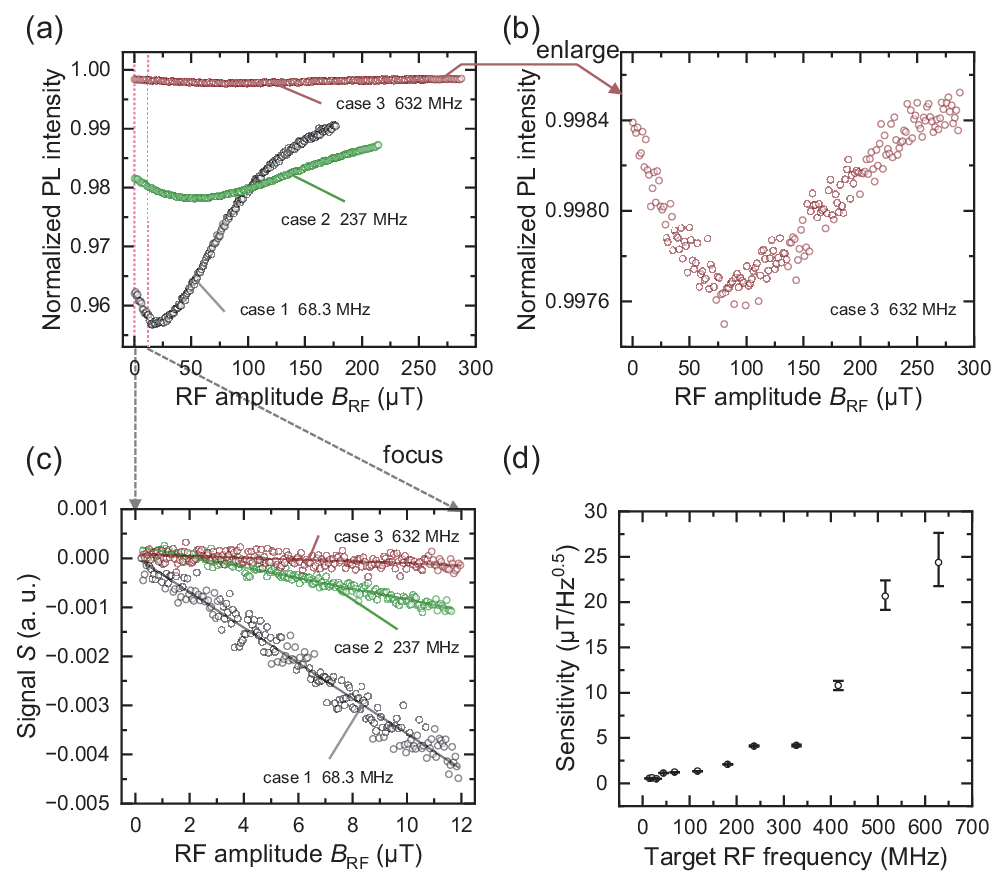}
\caption{(a) Changes in PL intensity induced by RF amplitude for cases 1--3. These measurements were performed under MW of $-20$ dBm. In all cases, the PL intensity initially decreases and then begins to increase above a threshold RF amplitude. (b) Magnified view of the PL intensity variation corresponding to case 3 in (a). (c) Variation in PL intensity (signal, $S$) up to approximately 12 $\mu$T for each case at a fixed MW frequency $f_{-0'}$. To enable a comparison of the changes in each case, the starting points of the plots are aligned. A decrease in PL intensity occurs. To estimate the RF sensitivity, $\mathrm{d}S/\mathrm{d}B_{\mathrm{RF}}$ was obtained by linear fitting ($S=aB_{\mathrm{RF}}+S_0$). (d) Sensitivities (\(\delta S/|\frac{\mathrm{d}S}{\mathrm{d}B_{\mathrm{RF}}}|\)) for various RF frequencies with error bar. The magnetic-field sensitivity degrades with increasing target frequency, as indicated by the increase in the noise-equivalent magnetic field. The error bars in (d) reflect the uncertainties from the linear fitting.}
\setlength\abovecaptionskip{3pt}
\label{signal}
\end{figure*}

Figure~\ref{signal}(a) and \ref{signal}(b) show dependence of the normalized PL intensity on $B_{\mathrm{RF}}$ for several RF frequencies (cases 1--3) at a fixed MW frequency $f_{-0'}$. At all frequencies, the normalized PL intensity nonmonotonically drops at weak RF amplitudes before rising at higher amplitudes. These dynamics are also apparent in the CW-ODMR spectra in Fig.~\ref{RFdependence}(a) (see the change in normalized PL intensity at \(f_{-0'}\) for black, blue, and red curves). We qualitatively attribute this nonmonotonic behavior to the competition between RF-induced population transfer and the spectral splitting of the RF-dressed states. When the RF Rabi frequency $\Omega_{\mathrm{RF}}$ is smaller than the linewidth, the transition between \(|+'\rangle\) and \(|-\rangle\) acts to reduce the steady-state population of $\vert{}0'\rangle$ (redistribution of the population). This effect leads to a decrease in the PL intensity. Conversely, when $\Omega_{\mathrm{RF}}$ exceeds the linewidth, well-resolved dressed states form, shifting and splitting the MW resonance and increasing MW detuning at $f_{-0'}$. This mechanism plays a role in recovering the PL intensity. A detailed analysis of this behavior is beyond the scope of the present work and will be reported elsewhere.

Based on the above behavior, we extracted \(\mathrm{d}S/\mathrm{d}B_{\mathrm{RF}}\) by linearly fitting the weak-RF regime, as shown in Fig.~\ref{signal}(c). Combining this with $\delta S$, the sensitivity across RF frequencies is ultimately given by \(\delta S/|\frac{\mathrm{d}S}{\mathrm{d}B_{\mathrm{RF}}}|\), as illustrated in Fig.~\ref{signal}(d). We have thus demonstrated CW-ODMR AC signal detection from 14.4~MHz to 632~MHz. While frequency limits currently stem from available transverse fields, stronger magnetic fields would further expand the frequency-tuning range. Notably, the sensitivity degrades at higher target frequencies. The optimal sensitivity is 0.558~$\mu{\rm T}/\sqrt{\rm Hz}$ at 14.4~MHz, whereas it degrades to 24.4~$\mu{\rm T}/\sqrt{\rm Hz}$ at 632~MHz. A higher RF frequency implies the application of a stronger transverse magnetic field. The underlying mechanisms tied to stronger bias fields are discussed in the following paragraphs.

We now discuss the physical mechanisms behind high-frequency sensitivity degradation. Although the reduced RF responsiveness $\cos\alpha$ in Eqs.~\eqref{eq:7}--\eqref{eq:10} contributes, this value remains high (0.983 at $B_x = 20\text{~mT}$, 0.965 at $B_x = 30\text{~mT}$, and 0.926 at $B_x = 50\text{~mT}$). Thus, this term does not explain the observed $\sim$44-fold sensitivity loss. We therefore consider the ODMR linewidths and contrasts under a strong transverse magnetic field to be the primary drivers of this degradation, and discuss strategies to improve sensitivity.

\begin{figure*}[ht]
\centering
\includegraphics[width=17cm]{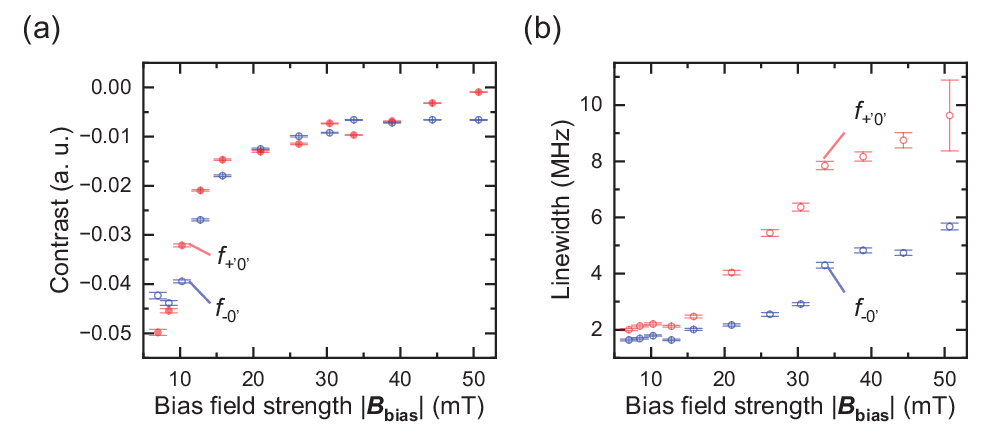}
\caption{(a) ODMR contrasts for $f_{+'0'}$ and $f_{-0'}$ transitions as a function of $\lvert \bm{B}_{\mathrm{bias}} \rvert$. The contrast is defined as the reduction from the baseline to the dip minimum obtained by Lorentzian fitting of the ODMR dip, taking a negative value. The red and blue dots correspond to $f_{+'0'}$ and $f_{-0'}$, respectively. In both cases, the contrast gradually decreases due to $B_x$-induced spin mixing. (b) CW-ODMR linewidths (FWHM) for the corresponding transitions as a function of $\lvert \bm{B}_{\mathrm{bias}} \rvert$. In both cases, the linewidth broadens with increasing $\lvert \bm{B}_{\mathrm{bias}} \rvert$, where the linewidth for the $f_{+'0'}$ transition is approximately twice that for $f_{-0'}$. All plots in (a) and (b) were measured at identical MW and laser powers.}
\setlength\abovecaptionskip{3pt}
\label{result5}
\end{figure*}

As shown in Fig.~\ref{result5}(a), the ODMR contrast for the $f_{+'0'}$ and $f_{-0'}$ transitions diminishes with increasing $\lvert \bm{B}_{\mathrm{bias}} \rvert$, exhibiting a sixfold reduction for the $f_{-0'}$ transition used in the sensitivity measurements. This stems from the $B_x\hat{S}_x$-induced spin mixing described by the corresponding term in Eq.~\eqref{eq:1}. Mixing between $\vert+\rangle$ and $\vert0\rangle$ alters the laser-driven photodynamics\cite{tetienne2012magnetic} by modifying the state lifetimes and suppressing transition selectivity during relaxation. Consequently, the PL contrast gradually decreases. 
Crucially, spin mixing is more pronounced in the optically excited states owing to their smaller zero-field splitting (approximately 1.42~GHz,\cite{doherty2013nitrogen} roughly half that of the optical ground states). The accessible frequency range will eventually be limited by the ODMR contrast becoming too small for practical detection. A rigorous quantitative evaluation of this $B_x$-dependent contrast requires the solution of steady-state rate equations in a seven-level model\cite{tetienne2012magnetic} using transition rates from time-resolved PL measurements, which is beyond the scope of this work.

Like the ODMR contrast, the linewidth directly determines the RF sensitivity. Figure~\ref{result5}(b) shows the $B_x$ dependence of the linewidths for the $f_{+'0'}$ and $f_{-0'}$ transitions. In both states, the linewidth broadens with increasing $B_x$.
For the $f_{-0'}$ transition, the linewidth at the maximum $B_x$ expands $\sim$3.5-fold relative to the minimum $B_x$. This change derives from altered noise susceptibility. While an on-axis field leaves the spin states primarily sensitive  to axial noise, strong perpendicular fields allow perpendicular fluctuations ($\delta B_x$) to contribute significantly to line broadening. Specifically, $\delta B_x$ causes a first-order shift in the MW transition frequencies $f_{+'0'}$ and $f_{-0'}$ (Eqs.~\eqref{eq:4} and \eqref{eq:5}). These slow fluctuations likely originate from the neodymium magnets and nearby P1 centers. Given the high nitrogen concentration of the sample ($1.4 \times 10^{19}~/\text{cm}^{3}$), the surrounding spin bath of the P1 centers are considered to provide a significant contribution to the observed broadening.

We also observe a clear linewidth asymmetry between $f_{+'0'}$ and $f_{-0'}$, with $f_{+'0'}$ exhibiting broader lines than $f_{-0'}$. From Eqs.~\eqref{eq:4} and \eqref{eq:5}, the first-order frequency shift induced by a small fluctuation $\delta B_x$ is approximately twice as large for $f_{+^\prime0^\prime}$ as for $f_{-0^\prime}$.\cite{qiu2021nuclear} This difference is consistent with the experimentally observed linewidth asymmetry. Evaluating sensitivity using the $f_{-0'}$ transition is therefore advantageous for performance optimization.

The loss of ODMR contrast and the increase in linewidth are consistent with reduced RF responsivity at larger transverse fields. These measurements identify contributions to the sensitivity degradation, but do not
constitute a complete quantitative decomposition of the measured noise-equivalent field. This issue is left for future work.

In the preceding paragraphs, we discussed the primary mechanisms that degrade the sensitivity at higher RF frequencies. To enhance sensitivity, several strategies can be implemented. One possible route is the use of an additional RF dressing field, which can narrow the effective ODMR linewidth and provide an additional degree of freedom to tune the dressed-state transition frequency. The generation of RF double-dressed states\cite{okaniwa2024frequency} causes the CW-ODMR linewidth to become narrow, thus increasing sensitivity. Furthermore, by accounting for the breakdown of the rotating wave approximation,\cite{okaniwa2024frequency, wang2021nanoscale} this approach could extend the detectable frequency range to approximately 1~GHz, which is about 1.5~times the accessible RF-frequency range demonstrated in this work. 

Second, spin-bath driving\cite{bauch2018ultralong} can mitigate $T_2^*$-limited dephasing in high-P1-density samples by continuously applying an AC field at a tuned frequency. While this approach is well-suited to samples with high concentrations of P1 centers, it requires precise tuning of the driving field frequency to interact with all surrounding P1 centers. We therefore need to research the full electron spin resonance spectrum of P1 centers under a transverse bias field. 

Finally, standard NV enhancement techniques--such as photon-collection optics (e.g., compound parabolic concentrators), lock-in detection, and optimizing $T_2^*$ and the NV concentrations--remain directly applicable for further performance gains in sensitivity.

In summary, we experimentally demonstrated CW-ODMR-based detection of AC magnetic fields from 14.4--632~MHz using RF-dressed NV electron-spin states under a strong transverse bias field. Unlike pulsed dynamical decoupling-based protocols, our method does not require high-power MW pulse sequences. The proposed method may be useful for RF-current and electromagnetic-field characterization in electronic devices, particularly when simple CW operation and spatially resolved measurements are desirable, while retaining potential compatibility with wide-field optical readouts. Nevertheless, CW-ODMR under strong perpendicular fields warrants further study, with emphasis on the linewidth evolution and noise sensitivity from weak to strong bias regimes. Our results provide a route toward simple, frequency-tunable NV magnetometry in the hundred-MHz regime.

See the supplementary material for details on quantum-mechanical calculations supporting our sensing scheme (Sec.~S1), experimental setup and procedure (Sec.~S2), and the methodology used to extract \(\{B_x, B_z\}\) from the experimentally obtained \(f_{+'0'}\) and \(f_{-0'}\) (Sec.~S3).

We are grateful to Dr. Ryo Tamaki for helpful discussions and suggestions that greatly improved our work.

This research is supported by the MEXT Quantum Leap Flagship Program (Q-LEAP) (Grant No. JPMXS0118067395), JSPS KAKENHI (Grant No. JP25K01292), JSPS Grant-in-Aid for JSPS Fellows (Grant No. JP24KJ1964), SIP Program, Program for the Advancement of
Next Generation Research Projects and CSRN, Keio University. This work was also supported by JST PRESTO (Grant No. JPMJPR245B), JST CREST (Grant Nos. JPMJCR24A5 and JPMJCR23I5), JST Moonshot R\&D (Grant No. JPMJMS226C), and JSPS Program for Forming Japan’s Peak Research Universities (J-PEAKS) (Grant No. JPJS00420230006). 

\section*{AUTHOR DECLARATIONS}

\subsection*{Conflict of Interest}
The authors have no conflicts to disclose.

\subsection*{Author Contributions}
{\bf Ryusei Okaniwa}: Conceptualization (lead); funding acquisition (supporting); investigation (lead); methodology (equal); project administration (supporting); software (supporting); visualization (lead); writing -- original draft (lead); writing -- review and editing (equal). {\bf Rui Suzuki}: Investigation (supporting); software (supporting); validation (supporting); writing -- review and editing (supporting). {\bf Yuichiro Matsuzaki}: Conceptualization (equal); data curation (supporting); formal analysis (lead); investigation (supporting); supervision (equal); validation (equal); writing -- review and editing (lead). {\bf Norio Tokuda}: Resources (equal); writing -- review and editing (supporting). {\bf Junko Ishi-Hayase}: Conceptualization (supporting); funding acquisition (lead); investigation (supporting); methodology (supporting); project administration (lead); resources (lead); supervision (lead); writing -- review and editing (lead).

\section*{Data Availability Statement}
The data that support the findings of this study are available from the corresponding authors upon reasonable request.

\nocite{*}
\section*{REFERENCES}
\bibliography{aipsamp}% Produces the bibliography via BibTeX.

\end{document}

\begin{center}

{\Large\bfseries Supplementary Material}

\vspace{3mm}

{\large\bfseries
Continuous-wave AC magnetometry up to hundreds of megahertz\\ using nitrogen-vacancy centers under a strong transverse magnetic field
}

\vspace{4mm}

Ryusei Okaniwa$^{1,2}$,
Rui Suzuki$^{1,2}$,
Yuichiro Matsuzaki$^{3}$,
Norio Tokuda$^{4}$,
Junko Ishi-Hayase$^{1,2,*}$

\vspace{3mm}

{\small

$^{1}$School of Fundamental Science and Technology, Keio University,
Yokohama, Kanagawa 223-8522, Japan

$^{2}$Center for Spintronics Research Network, Keio University,
Yokohama, Kanagawa 223-8522, Japan

$^{3}$Department of Electrical, Electronic, and Communication Engineering,
Faculty of Advanced Science and Engineering,
Chuo University,
Kasuga, Bunkyo-ku, Tokyo 112-8551, Japan

$^{4}$Advanced Research Center for Diamond Science and Technology,
Kanazawa University,
Kanazawa, Ishikawa 920-1192, Japan

$^{*}$Corresponding author:
hayase@appi.keio.ac.jp 

}

\end{center}

%%%%%%%%%%%%%%%%%%%%%%%%%%%%%%%%%%%%%%%%%%%%%%%%%%%%%%%%%%%%
\section{Detailed theory under a strong transverse magnetic field}
\label{sec:1}
In this section, we present the theoretical details and calculation procedures underlying the results described in the main text. We focus on the regime in which a strong transverse magnetic field \(B_x\) is applied to the nitrogen-vacancy (NV) centers. 
 
As described in the main text, we begin with the total Hamiltonian \(\hat{H}_0\):
\begin{align}
\label{eq:1}
\hat{H}_0=D\hat{S}_z^2 + E_x(\hat{S}_x^2 -\hat{S}_y^2)+\gamma_eB_x\hat{S}_x,
\end{align}
where \(\hat{S}_i\) (\(i = x, y, z\)) is the spin-1 operator of the NV electron spin, \(D/2\pi\) (\(\simeq2.87\)~GHz) is the zero-field splitting, \(E_x/2\pi\) is the intrinsic strain along the \(x\) direction, and \(\gamma_e/2\pi\) (\(\simeq28\)~GHz/T) is the gyromagnetic ratio of the electron spin. 
Equation ~\eqref{eq:1} can be exactly diagonalized, allowing \(\hat{H}_0\) to be written as follows:
\begin{align}
\label{eq:2}
\hat{H}_0=E_{+'}|+'\rangle\langle +'|+(D-E_x)|-\rangle\langle -|+E_{0'}|0'\rangle\langle 0'|,
\end{align}
where \(E_{+'}\) and \(E_{0'}\) are given by 
\begin{align}
\label{eq:3}
E_{+'}=\frac{1}{2}\{(D+E_x)+\sqrt{(D+E_x)^2+(2\gamma_e B_x)^2}\},
\end{align}
\begin{align}
\label{eq:4}
E_{0'}=\frac{1}{2}\{(D+E_x)-\sqrt{(D+E_x)^2+(2\gamma_e B_x)^2}\}.
\end{align}
The new eigenstates \(|+'\rangle\) and \(|0'\rangle\) are superpositions of \(|+\rangle\) and \(|0\rangle\). Specifically,  \(|+\rangle\) and \(|0\rangle\) are expressed using \(|+'\rangle\) and \(|0'\rangle\) as follows \cite{matsuzaki2025theoretical}:
\begin{align}
\label{eq:5}
|+\rangle=\cos\alpha |+'\rangle-\sin\alpha |0'\rangle,\ |0\rangle=\sin\alpha |+'\rangle+\cos\alpha |0'\rangle.
\end{align}
Here, \(\alpha\) satisfies the following two equations: 
\begin{equation}
\begin{aligned}
\cos({2\alpha})&=(D+E_x)/\sqrt{(D+E_x)^2+(2\gamma_e B_x)^2},\\
\sin({2\alpha})&=2\gamma_eB_x/\sqrt{(D+E_x)^2+(2\gamma_e B_x)^2}.
\end{aligned}
\label{eq:6}
\end{equation}
As the bias field strength increases, this mixing effect becomes more pronounced. The transition microwave (MW) frequencies \(f_{+'0'}\) (\(|+'\rangle \Leftrightarrow |0'\rangle\))  and \(f_{-0'}\) (\(|-\rangle \Leftrightarrow |0'\rangle\)) without the driving radiofrequency (RF) magnetic field can be written as
\begin{eqnarray}
\label{eq:7-8}
2\pi f_{+'0'}&&=E_{+'}-E_{0'}=\sqrt{(D+E_x)^2+(2\gamma_eB_x)^2},\\
2\pi f_{-0'}&&=(D-E_x)-E_{0'}=\frac{1}{2}D-\frac{3}{2}E_x+\frac{1}{2}\sqrt{(D+E_x)^2+(2\gamma_eB_x)^2}.
\end{eqnarray}
In continuous-wave optically detected magnetic resonance (CW-ODMR) measurements, a photoluminescence (PL) dip is expected to be observed at  \(f_{+'0'}\) and \(f_{-0'}\).  Both resonant frequencies shift toward higher frequencies \cite{beaver2024optimizing} and the energy difference \(E_{+'-}\) between \(|+'\rangle\) and \(|-\rangle\) (equal to \(E_{+'}-E_{-}=2\pi(f_{+'0'}-f_{-0'})\)) increases rapidly according to:
\begin{align*}
\label{eq:9}
E_{+'-}&=E_{+'}-E_{-}=2\pi(f_{+'0'}-f_{-0'})\\
&=-\frac{1}{2}D+\frac{3}{2}E_x+\frac{1}{2}\sqrt{(D+E_x)^2+(2\gamma_eB_x)^2}.
\stepcounter{equation}\tag{\theequation}
\end{align*}

Next, we include the MW and RF (target) fields in addition to the diagonalized Hamiltonian \(\hat{H}_0\). The subsequent calculations provide the essential framework for understanding the formation of the RF-dressed states. The total Hamiltonian \(\hat{H}_{\mathrm{total}}\) including the above driving fields is expressed as
\begin{align*}
\label{eq:10}
\hat{H}_{\mathrm{total}}=&\hat{H}_0+\hat{H}_{\mathrm{drive}}(t)\\
=&\hat{H}_0 + \gamma_e B_{\mathrm{RF}}\hat{S}_z\cos(\omega_{\mathrm{RF}}t)\\
&+\gamma_e B_{\mathrm{MW}x}\hat{S}_x\cos(\omega_{\mathrm{MW}}t),
\stepcounter{equation}\tag{\theequation}
\end{align*}
where \(\hat{H}_{\mathrm{drive}}(t)\) is the time-dependent Hamiltonian of the MW and RF field, \(B_{\mathrm{RF}}\) (\(B_{\mathrm{MW}x}\)) is the amplitude of the RF field (MW), and \(\omega_{\mathrm{RF}}\) (\(\omega_{\mathrm{MW}}\)) is the angular frequency of the RF field (MW). For simplicity, only the \(x\)-component of the MW is considered. The explicit form of the Hamiltonian for the \(y\)-component of the MW is provided below. We also assume that
the target RF field is polarized parallel to the NV axis. 

To remove the time-dependent term of MW in Eq.~\eqref{eq:10}, we adopt a rotating frame with \(\hat{U}^{(1)}=\exp[i\omega_{\mathrm{MW}}(|+'\rangle\langle +'|+|-\rangle\langle -|)t]\). In the basis of the eigenstates \{\(|+'\rangle, |-\rangle, |0'\rangle\)\},  the effective Hamiltonian \(\hat{H}^{(1)}_{\mathrm{total}}\) after the rotating wave approximation (RWA) is given by
\begin{align*}
\label{eq:11}
\hat{H}^{(1)}_{\mathrm{total}}=&\hat{U}^{(1)}\hat{H}_{\mathrm{total}}\hat{U}^{(1)\dagger}-i\hat{U}^{(1)}\frac{\mathrm{d}}{\mathrm{d}t}\hat{U}^{(1)\dagger}\\
\simeq &(E_{+'}-\omega_{\mathrm{MW}})|+'\rangle\langle +'|\\
&+(D-E_x-\omega_{\mathrm{MW}})|-\rangle\langle -|+E_{0'}|0'\rangle\langle 0'|\\
&+\gamma_e B_{\mathrm{RF}}\cos \alpha(|+'\rangle\langle -| + \mathrm{h.\ c.})\cos(\omega_{\mathrm{RF}}t)\\
&+\frac{\gamma_e B_{\mathrm{MW}x}}{2}(\cos^2\alpha-\sin^2\alpha)(|+'\rangle\langle 0'|+\mathrm{h.\ c.}).
\stepcounter{equation}\tag{\theequation}
\end{align*}
Furthermore, we define another rotating frame as \(\hat{U}^{(2)}=\exp[i\frac{\omega_{\mathrm{RF}}}{2}(|+'\rangle\langle +'|-|-\rangle\langle -|)t]\) and apply this to the Hamiltonian \(\hat{H}^{(1)}_{\mathrm{total}}\) in Eq.~\eqref{eq:11} under the RWA. The new effective Hamiltonian \(\hat{H}^{(2)}_{\mathrm{total}}\) after the RWA is given by
\begin{align*}
\label{eq:12}
\hat{H}^{(2)}_{\mathrm{total}}=&\hat{U}^{(2)}\hat{H}^{(1)}_{\mathrm{total}}\hat{U}^{(2)\dagger}-i\hat{U}^{(2)}\frac{\mathrm{d}}{\mathrm{d}t}\hat{U}^{(2)\dagger}\\
\simeq &\hat{H}^{(2)}_{\mathrm{total}0} + \hat{H}^{(2)}_{\mathrm{total}}(t),
\stepcounter{equation}\tag{\theequation}
\end{align*}
\begin{align*}
\label{eq:13}
\hat{H}^{(2)}_{\mathrm{total}0}=&(E_{+'}-\omega_{\mathrm{MW}}-\frac{\omega_{\mathrm{RF}}}{2})|+'\rangle\langle +'|\\
&+(D-E_x-\omega_{\mathrm{MW}}+\frac{\omega_{\mathrm{RF}}}{2})|-\rangle\langle -|+E_{0'}|0'\rangle\langle 0'|\\
&+\frac{\gamma_e B_{\mathrm{RF}}}{2}\cos \alpha(|+'\rangle\langle -| + \mathrm{h.\ c.}),
\stepcounter{equation}\tag{\theequation}
\end{align*}
\begin{align}
\label{eq:14}
\hat{H}^{(2)}_{\mathrm{total}}(t)=\frac{\gamma_e B_{\mathrm{MW}x}}{2}(\cos^2\alpha-\sin^2\alpha)(e^{i\frac{\omega_{\mathrm{RF}}}{2}t}|+'\rangle\langle 0'|+\mathrm{h.\ c.}).
\end{align}
According to Eqs.~\eqref{eq:12}--\eqref{eq:14}, the Hamiltonian \(\hat{H}^{(2)}_{\mathrm{total}}\) can be divided into a time-independent part \(\hat{H}^{(2)}_{\mathrm{total}0}\) and a time-dependent part \(\hat{H}^{(2)}_{\mathrm{total}}(t)\). Here, we move to the interaction picture. The time-independent Hamiltonian \(\hat{H}^{(2)}_{\mathrm{total}0}\) can be diagonalized as follows:
\begin{align}
\label{eq:15}
\hat{H}^{(2)}_{\mathrm{total}0}=E_{\widetilde{+'}}
|\widetilde{+'}\rangle\langle \widetilde{+'}|+E_{\widetilde{-}}|\widetilde{-}\rangle\langle \widetilde{-}|+E_{0'}|0'\rangle\langle 0'|,
\end{align}
where the new eigenlevels of \(E_{\widetilde{+'}}\) and \(E_{\widetilde{-}}\) are given by
\begin{align*}
\label{eq:16}
E_{\widetilde{+'}}=&\frac{1}{2}[(E_{+'}+D-E_x-2\omega_{\mathrm{MW}})\\
&+\sqrt{(E_{+'}-D+E_x-\omega_{\mathrm{RF}})^2+(\gamma_e B_{\mathrm{RF}}\cos\alpha)^2}],
\stepcounter{equation}\tag{\theequation}
\end{align*}
\begin{align*}
\label{eq:17}
E_{\widetilde{-}}=&\frac{1}{2}[(E_{+'}+D-E_x-2\omega_{\mathrm{MW}})\\
&-\sqrt{(E_{+'}-D+E_x-\omega_{\mathrm{RF}})^2+(\gamma_e B_{\mathrm{RF}}\cos\alpha)^2}].
\stepcounter{equation}\tag{\theequation}
\end{align*}
The new eigenstates \(|\widetilde{+'}\rangle\) and \(|\widetilde{-}\rangle\) are superpositions of \(|+'\rangle\) and \(|-\rangle\). Therefore, \(|+'\rangle\) and \(|-\rangle\) are written using \(|\widetilde{+'}\rangle\) and \(|\widetilde{-}\rangle\) as follows \cite{matsuzaki2025theoretical}:
\begin{align}
\label{eq:18}
|+'\rangle=\cos\beta |\widetilde{+'}\rangle-\sin\beta |\widetilde{-}\rangle,\ |-\rangle=\sin\beta |\widetilde{+'}\rangle+\cos\beta |\widetilde{-}\rangle,
\end{align}
where \(\beta\) is given by 
\begin{equation}
\begin{aligned}
\cos({2\beta})&=(E_{+'}-D+E_x-\omega_{\mathrm{RF}})/\sqrt{(E_{+'}-D+E_x-\omega_{\mathrm{RF}})^2+(\gamma_e B_{\mathrm{RF}}\cos\alpha)^2},\\
\sin(2\beta)&=\gamma_e B_{\mathrm{RF}}\cos\alpha/\sqrt{(E_{+'}-D+E_x-\omega_{\mathrm{RF}})^2+(\gamma_e B_{\mathrm{RF}}\cos\alpha)^2}.
\end{aligned}
\label{eq:19}
\end{equation}
Rewriting the Hamiltonian \(\hat{H}^{(2)}_{\mathrm{total}}(t)\) in Eq.~\eqref{eq:14} using the new basis defined by Eq.~\eqref{eq:18}, the Hamiltonian \(\hat{H}^{(2)}_{\mathrm{total}I}\) in the interaction picture is given by
\begin{align*}
\label{eq:20}
\hat{H}^{(2)}_{\mathrm{total}I}=&e^{i\hat{H}^{(2)}_{\mathrm{total}0}t}\hat{H}^{(2)}_{\mathrm{total}}(t)e^{-i\hat{H}^{(2)}_{\mathrm{total}0}t}\\
=&\frac{\gamma_e B_{\mathrm{MW}x}}{2}(\cos^2\alpha-\sin^2\alpha)\\
&\times[e^{i(E_{\widetilde{+'}}-E_{0'}+\frac{\omega_{\mathrm{RF}}}{2})t}\cos\beta|\widetilde{+'}\rangle\langle 0'|\\
&-e^{i(E_{\widetilde{-}}-E_{0'}+\frac{\omega_{\mathrm{RF}}}{2})t}\sin\beta|\widetilde{-}\rangle\langle 0'|+ \mathrm{h.\ c.}].
\stepcounter{equation}\tag{\theequation}
\end{align*}
Using Fermi’s golden rule \cite{wang2015strongly}, resonance occurs at \(\omega_{\mathrm{MW}}\) satisfying either Eq.~\eqref{eq:21} or Eq.~\eqref{eq:22}, where the observation of the RF-dressed states is expected from the CW-ODMR experiments.
\begin{align}
\label{eq:21}
E_{\widetilde{+'}}-E_{0'}+\frac{\omega_{\mathrm{RF}}}{2}=0,
\end{align}
\begin{align}
\label{eq:22}
E_{\widetilde{-}}-E_{0'}+\frac{\omega_{\mathrm{RF}}}{2}=0.
\end{align}
In this study, we focus on the case where the RF field is resonant with the energy difference between \(|+'\rangle\) and \(|-\rangle\), i.e., \(\omega_{\mathrm{RF}}=E_{+'}-(D-E_x)\). Therefore, Eqs.~\eqref{eq:21} and \eqref{eq:22} can be further simplified, and the corresponding resonance frequencies (\(f_{\mathrm{MWresA}}\) and \(f_{\mathrm{MWresB}}\)) are obtained as follows:
\begin{align}
\label{eq:23}
2\pi f_{\mathrm{MWresA}}=\sqrt{(D+E_x)^2+(2\gamma_eB_x)^2}+\frac{\gamma_eB_{\mathrm{RF}}}{2}\cos\alpha,
\end{align}
\begin{align}
\label{eq:24}
2\pi f_{\mathrm{MWresB}}=\sqrt{(D+E_x)^2+(2\gamma_eB_x)^2}-\frac{\gamma_eB_{\mathrm{RF}}}{2}\cos\alpha.
\end{align}

Following a similar derivation, two resonance frequencies (\(f_{\mathrm{MWresC}}\) and \(f_{\mathrm{MWresD}}\)) can be obtained for the case of the \(y\)-polarized MW:
\begin{align}
\label{eq:25}
2\pi f_{\mathrm{MWresC}}=\frac{1}{2}D-\frac{3}{2}E_x+\frac{1}{2}\sqrt{(D+E_x)^2+(2\gamma_eB_x)^2}+\frac{\gamma_eB_{\mathrm{RF}}}{2}\cos\alpha,
\end{align}
\begin{align}
\label{eq:26}
2\pi f_{\mathrm{MWresD}}=\frac{1}{2}D-\frac{3}{2}E_x+\frac{1}{2}\sqrt{(D+E_x)^2+(2\gamma_eB_x)^2}-\frac{\gamma_eB_{\mathrm{RF}}}{2}\cos\alpha.
\end{align}

Equations \eqref{eq:23}--\eqref{eq:26} are also presented in the main text. Based on this theoretical framework, resonant RF radiation induces a coherent interaction between $\vert{}+'\rangle$ and $\vert{}-\rangle$, splitting each energy level and yielding four distinct levels corresponding to Eqs.~\eqref{eq:23}--\eqref{eq:26}. In CW-ODMR measurements, four PL dips are expected to appear. According to Eqs.~\eqref{eq:23}--\eqref{eq:24} and Eqs.~\eqref{eq:25}--\eqref{eq:26}, the splitting between adjacent energy levels increases in proportion to \(B_{\text{RF}}\). The proportionality constant of this splitting depends on \(\cos\alpha\). The target RF field can be detected through the change in PL at a fixed MW frequency using CW-ODMR.
%%%%%%%%%%%%%%%%%%%%%%%%%%%%%%%%%%%%%%%%%%%%%%%%%%%%%%%%%%%%

\section{Experimental setup and procedure}
\begin{figure}[ht]
\centering
\includegraphics[width=15cm]{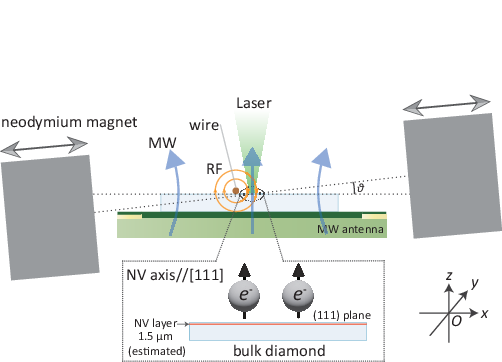}
\caption{Experimental setup around the diamond sample. A green laser beam was focused onto the near-surface NV layer (its thickness was estimated to be 1.5~\(\mu\)m). MW fields were supplied from bottom-mounted antenna and the RF fields (target fields) were applied via a 30~\(\mu\)m-copper wire. Ideally, the NV axis should be oriented along the $z$-direction in the laboratory frame. For precise magnetic field alignment, the neodymium magnets were mounted on motorized stages and fixtures to adjust the azimuthal angle ($\phi$) and polar angle ($\theta$). These motorized stages and fixtures were designed such that their center of rotation coincided with the diamond sample. The distance between the magnets and the diamond sample could be adjusted using actuators.}
\setlength\abovecaptionskip{3pt}
\label{S1:setup}
\end{figure}

The specific experimental setup is shown in Fig.~\ref{S1:setup}. We used a home-made confocal microscopic system and a bulk diamond sample, with an estimated NV center density of \(10^{16}\text{~/cm}^{3}\). A 532-nm Gaussian beam was focused onto the ensemble NV centers through a dry objective lens with a numerical aperture of 0.9 (Nikon, LU Plan Fluor 100X). In this experiment, the laser power was fixed at about $50\text{~}\mu\text{W}$ directly above the objective lens. The measurement point was set approximately $15\text{~}\mu\text{m}$ from the edge of the copper wire, with the positioning achieved using a three-axis piezo stage (Piezosystem Jena, TRITOR 102 SG) connected to the objective lens. The PL from the NV centers was collected by the same objective lens and detected by an avalanche photodiode (Excelitas, SPCM-AQRH-14-FC). The MW signals were generated by a vector signal generator (Rohde \& Schwarz, SMBV100A) and supplied to a dedicated antenna via a high-gain amplifier (Mini-Circuits, ZHL-16W-43-S+). During the acquisition of CW-ODMR spectra, the MW power was set to $-30\text{~dBm}$. Under some of the stronger magnetic fields, ODMR measurements were performed at a MW power of $-20\text{~dBm}$, rather than $-30\text{~dBm}$, to improve the contrast. The RF field was output from another vector signal generator. To obtain sufficient RF power at high frequencies, an amplifier (Keysight Technologies, 8347A and Mini-Circuits, LZY-22+) was placed after the generator. The amplified RF wave was delivered to the copper wire running across the diamond, applying an AC magnetic field to the NV centers. The opposite end of the copper wire was connected in parallel to a $50\text{~}\Omega$ terminator and an oscilloscope (RIGOL, MHO98). This configuration accounted for the frequency-dependent characteristics of both the generator and the amplifier. The RF amplitude was calibrated based on the Zeeman splitting of NV centers under a parallel magnetic field in combination with the frequency response derived from this setup. The bias magnetic field was applied by a pair of cylindrical neodymium magnets (diameter 30~mm, height 25~mm). Precise adjustment of these magnets is crucial for ensuring that the magnetic field is applied perpendicular to the NV axis. To address this, custom jigs and motorized stages were used to precisely align the magnetic field. Furthermore, the magnetic field intensity was made tunable by adjusting the distance from the diamond substrate using motorized actuators connected to each neodymium magnet.

Based on the above setup, we performed CW-ODMR-based AC magnetometry. To suppress laser fluctuations, the photon counts with the MW turned on and off were recorded separately to normalize the PL intensity. The photon-number fluctuations were evaluated as the standard deviation of the normalized photon counts. The photon counts with the MW on and off were acquired for equal durations and normalized by taking their ratio, with the MW-on photon count divided by the MW-off photon count. By repeating this procedure for various photon acquisition times, the normalized fluctuations were measured and plotted on a log-log scale. A linear fit on the log-log plot was used to determine the photon-number fluctuation at an acquisition time of 1~s ($\delta S$). CW-ODMR spectra were recorded while varying the distance between the motorized actuator and the diamond substrate. By narrowing the MW frequency sweep range to the vicinity of the resonance frequencies, the frequency of the applied RF could be estimated more precisely. Subsequently, the RF power at the estimated frequency was calibrated using the oscilloscope, and CW-ODMR spectra under the RF application were recorded. Finally, the MW frequency was fixed at $f_{-0'}$, and the variations in the normalized PL intensity (signal $S$) were recorded as a function of the RF amplitude.

%%%%%%%%%%%%%%%%%%%%%%%%%%%%%%%%%%%%%%%%%%%%%%%%%%%%%%%%%%%%

\section{Extracting information about the bias magnetic field}
In this section, we describe the method used to determine each component of the bias magnetic field $\bm{B}_{\mathrm{bias}}$. In the main text and Sec.~\ref{sec:1}, we focus exclusively on the off-axis component of the static field ($B_x$). Experimentally, however, a magnetic field component is also present along the NV axis ($B_z$). This arises from the misalignment of the neodymium magnets and the contribution of the Earth's magnetic field. Therefore, an additional term is added to Eq.~(\ref{eq:1}). The new Hamiltonian $\hat{H}_{0'}$ can then be written as follows:
\begin{align}
\label{eq:27}
\hat{H}_{0'} = D\hat{S}_z^2 + E_x(\hat{S}_x^2 - \hat{S}_y^2) + E_y(\hat{S}_x\hat{S}_y + \hat{S}_y\hat{S}_x) + \gamma_e B_x\hat{S}_x + \gamma_e B_z\hat{S}_z.
\end{align}
Here, we have included not only the strain in the $x$-direction ($E_x$), but also that in the $y$-direction ($E_y$). In the case of $B_x = B_z = 0$, the Hamiltonian $\hat{H}_{0'}$ can be exactly diagonalized, and each eigenstate can be denoted as $\{|+\rangle, |-\rangle, |0\rangle\}$. Under this condition, the energy difference between $|+\rangle$ and $|-\rangle$ ($E_{0}$) is expressed as $2\sqrt{E_x^2 + E_y^2}$. We experimentally evaluated $E_0/2\pi $ by performing CW-ODMR measurements at zero magnetic field, yielding $E_0/2\pi \simeq 4.13~\text{MHz}$. Individual components of the strain field cannot be extracted independently using only the CW-ODMR spectrum.

To determine $B_x$, $B_z$, and $|\bm{B}_{\mathrm{bias}}|$ from the two experimentally measured resonant frequencies ($f_{+'0'}$ and $f_{-0'}$), we employed a gradient descent algorithm based on the Hamiltonian $\hat{H}_{0'}$. The zero-field splitting was determined to be $D/2\pi = 2.87065~\text{GHz}$ from the ODMR spectrum. Moreover, because the individual values of $E_x$ and $E_y$ cannot be experimentally isolated, 1000 sets of strain parameters were generated following a distribution with zero mean and a standard deviation of $E_x^2 + E_y^2$, $\sigma = 1.65~\text{MHz}$ \cite{matsuzaki2016optically}. These mean value and standard deviation reproduced the experimental value of $E_0 / 2\pi = 4.13\text{ MHz}$ obtained at zero magnetic field. For a given set of $\{D, E_x, E_y\}$, the field components $\{B_x, B_z\}$ that minimize the discrepancy between the experimental resonance frequencies ($f_{+'0'}$ and $f_{-0'}$) and the calculated frequencies ($f_{\text{cal}+'0'}$ and $f_{\text{cal}-0'}$) derived from the Hamiltonian in Eq.~\eqref{eq:27} were determined via gradient descent within an appropriate search range \cite{matsuzaki2016optically}. Based on the above, the cost function \(\text{cost}(B_x, B_z)\) is specifically expressed as
\begin{align}
\text{cost}(B_x, B_z)&=  X^2(B_x,B_z)+Y^2(B_x,B_z),\label{eq:28}\\
X(B_x,B_z)&=f_{+'0'}-f_{\text{cal}+'0'}(B_x,B_z),\label{eq:29}\\
Y(B_x,B_z)&=f_{-0'}-f_{\text{cal}-0'}(B_x,B_z).\label{eq:30}
\end{align}

Table \ref{tabS1} presents the optimized values of $B_x$, $B_z$, and $|\bm{B}_{\mathrm{bias}}|$ for each experimentally obtained value $E_{+'-}/2\pi$ ($= f_{+'0'} - f_{-0'}$). The search ranges for $\{B_x, B_z\}$ at each value of $E_{+'-}/2\pi$ are listed in Table \ref{tabS2}. In all cases, the values of 
$B_x$ and $B_z$ converged sufficiently, and the resulting residuals were approximately five orders of magnitude smaller than the ODMR linewidth, expressed in MHz (see the ``residuals (MHz)'' column in Table \ref{tabS2}). Furthermore, the perpendicular component $B_x$ was found to account for more than 99.5\% of the total magnitude of the static magnetic field. The increase in the relative contribution of \(B_z\) with increasing $E_{+'-}/2\pi$ is likely to be caused by greater misalignment as the neodymium magnets were brought closer to the diamond. The horizontal axis of Fig.~2(a) in the main text represents $|\bm{B}_{\mathrm{bias}}|$ determined using the procedure described above.
\begin{table}[htbp]
    \centering
    \caption{Values of $\{B_x, B_z\}$ for each $E_{+'-}/2\pi$, determined using gradient descent implemented in Wolfram Mathematica. The symmetry of the Hamiltonian in Eq.~\eqref{eq:27} means that the sign of $B_z$ cannot be uniquely determined from the ODMR spectrum; therefore, only its absolute value $\vert{}B_z\vert{}$ is reported. The magnitude $\vert{}\bm{B}_{\mathrm{bias}}\vert{}$ was subsequently calculated from the obtained set of $\{B_x, B_z\}$.}
    \label{tabS1}
    \begin{tabular}{cccc}
        \hline
        $E_{+'-}/2\pi$ (MHz) & $B_x$ (mT) & $B_z$ (mT) & $|\bm{B_{\mathrm{bias}}}|$ (mT) \\
        \hline
        14.4 & 7.01 & 0.0806 &  7.01  \\ 
        20.2 & 8.50 & 0.0776 &  8.50  \\
        29.6 & 10.3 & 0.104 &  10.3  \\
        44.4 & 12.8 & 0.0805 &  12.8  \\
        68.3 & 15.8 & 0.252 &  15.8  \\
        116 & 21.0& 0.232 & 21.0  \\
        181 & 26.2 & 0.690 & 26.2  \\
        237 & 30.4 & 0.742 & 30.4 \\
        327 & 33.5 & 3.16 & 33.7 \\
        416 & 38.7 & 3.80 & 38.9 \\
        516 & 44.2 & 4.46 & 44.4 \\
        632 & 50.4 & 4.98 & 50.7 \\
    \end{tabular}
\end{table}

\begin{table}[htbp]
    \centering
    \caption{Search ranges of $\{B_x, B_z\}$ and corresponding residuals $\vert{}X(B_x, B_z)\vert{}$ and $\vert{}Y(B_x, B_z)\vert{}$ for each value of $E_{+'-}/2\pi$ during gradient descent optimization. The absolute values of the residuals are reported.}
    \label{tabS2}
    \begin{tabular}{ccc}
        \hline
        $E_{+'-}/2\pi$ (MHz) & search ranges of \(\{B_x,B_z\}\) (mT) &  residuals (MHz)\\
        \hline
        14.4 & $0\leq B_x\leq50, 0\leq B_z\leq5$ & $|X(B_x,B_z)|=7.48\times10^{-7}, |Y(B_x,B_z)|=4.10\times10^{-7}$  \\
        20.2 & $0\leq B_x\leq50, 0\leq B_z\leq5$ & $|X(B_x,B_z)|=3.74\times10^{-6}, |Y(B_x,B_z)|=7.09\times10^{-6}$ \\
        29.6 & $0\leq B_x\leq50, 0\leq B_z\leq5$ & $|X(B_x,B_z)|=1.01\times10^{-6}, |Y(B_x,B_z)|=3.11\times10^{-6}$ \\
        44.4 & $0\leq B_x\leq50, 0\leq B_z\leq5$ & $|X(B_x,B_z)|=2.47\times10^{-6}, |Y(B_x,B_z)|=8.67\times10^{-6}$ \\
        68.3 & $0\leq B_x\leq50, 0\leq B_z\leq5$ & $|X(B_x,B_z)|=1.71\times10^{-5}, |Y(B_x,B_z)|=1.16\times10^{-5}$ \\
        116 & $0\leq B_x\leq50, 0\leq B_z\leq5$ & $|X(B_x,B_z)|=9.14\times10^{-7}, |Y(B_x,B_z)|=2.33\times10^{-6}$ \\
        181 &$0\leq B_x\leq50, 0\leq B_z\leq5$ &  $|X(B_x,B_z)|=2.49\times10^{-6}, |Y(B_x,B_z)|=7.98\times10^{-6}$   \\
        237 & $0\leq B_x\leq50, 0\leq B_z\leq5$  & $|X(B_x,B_z)|=4.81\times10^{-6}, |Y(B_x,B_z)|=8.00\times10^{-7}$ \\
        327 & $0\leq B_x\leq50, 0\leq B_z\leq5$  &  $|X(B_x,B_z)|=1.54\times10^{-5}, |Y(B_x,B_z)|=4.15\times10^{-6}$  \\
        416 & $0\leq B_x\leq50, 0\leq B_z\leq5$ & $|X(B_x,B_z)|=7.35\times10^{-7}, |Y(B_x,B_z)|=1.26\times10^{-6}$  \\
        516 & $0\leq B_x\leq50, 0\leq B_z\leq5$ & $|X(B_x,B_z)|=6.45\times10^{-6}, |Y(B_x,B_z)|=8.02\times10^{-6}$  \\
        632 & $30\leq B_x\leq70, 3\leq B_z\leq7$ & $|X(B_x,B_z)|=7.86\times10^{-7}, |Y(B_x,B_z)|=2.58\times10^{-6}$\\
    \end{tabular}
\end{table}

%%%%%%%%%%%%%%%%%%%%%%%%%%%%%%%%%%%%%%%%%%%%%%%%%%%%%%%%%%%%
\clearpage
\bibliographystyle{apsrev4-2}
\bibliography{SI_reference}